\documentclass[nofootinbib,reprint,amsmath,amssymb,aps]{revtex4-1}
\usepackage{graphicx}
\usepackage[utf8,latin1]{inputenc}
\usepackage{dcolumn}
\usepackage{bm}
\usepackage{mathrsfs}
\usepackage{hyperref}
\usepackage{amsmath}
\usepackage{longtable}
\usepackage{multirow}
\newcommand{\qm}[1]{``#1''}
\usepackage{hyperref}
\hypersetup{colorlinks, linkcolor={red},citecolor={blue},urlcolor={blue}}  
\newcommand\ChangeRT[1]{\noalign{\hrule height #1}}

\begin{document}

\title[Timescales for the onset of chaos in the general relativistic Poynting-Robertson effect]{Timescales of the chaos onset in the\\ general relativistic Poynting-Robertson effect}

\author{Vittorio De Falco$^{1}$}\email{vittorio.defalco@physics.cz}
\author{William Borrelli$^{2}$} \email{william.borrelli@unicatt.it}

\affiliation{$^1$Department of Mathematics and Applications \qm{R. Caccioppoli}, University of Naples Federico II, Via Cintia, 80126 Naples, Italy, \\
$^2$Dipartimento di Matematica e Fisica, Universit\`a Cattolica del Sacro Cuore, Via dei Musei 41, 25121, Brescia, Italy.}

\date{\today}

\begin{abstract}
It has been proved that the general relativistic Poynting-Robertson effect in the equatorial plane of Kerr metric shows a chaotic behavior for a suitable range of parameters. As a further step, we calculate the timescale for the onset of chaos through the Lyapunov exponents, estimating how this trend impacts on the observational dynamics. We conclude our analyses with a discussion on the possibility to observe this phenomenon in neutron star and black hole astrophysical sources.
\end{abstract}

\maketitle

\section{Introduction}
\label{sec:intro}
The general relativistic Poynting-Robertson (PR) effect is a phenomenon occurring in high-energy astrophysics. The motion of accreting matter in the vicinity of compact objects, such as black holes (BHs) and neutron stars (NSs), is strongly affected by the gravitational field. In addition, there could be the presence of other perturbing effects responsible to alterate the geodesic motion of the surrounding matter (e.g., magnetic fields, hydrodynamical/magneto-hydrodynamical processes). 

The X-ray electromagnetic radiation produced by an emitting source located close to the compact object (e.g., very hot corona around supermassive BHs in AGNs, stellar mass BHs, and weakly magnetised NSs in X-ray binary systems \citep{Fabian2015,Reis2013,Done2007}, boundary layers around NSs or the NS surface itself \citep{Inogamov1999,Popham2001}, thermonuclear flashes occurring in the outermost layers of accreting NSs \citep{Lewin1993}) can intercept the inflowing matter modifying its motion. Indeed, the radiation force acting on relatively small-sized matter elements, treated as point-like particles, is composed by the radiation pressure (being opposite to the gravitational attraction) and the PR radiation drag force, originating when the test particle absorbs the incoming radiation and then re-emit it isotropically in its rest frame \cite{Poynting1903,Robertson1937,Bini2009,Bini2011}. The PR effect removes very efficiently energy and angular momentum from the affected test particle, configuring thus as a \emph{dissipative and non-linear dynamical system in GR}. 

From a modelling perspective, there are different treatments of the general relativistic PR effect going from Schawarzschild and Kerr to also other spacetimes from the two dimensional (2D) \cite{Bini2009,Bini2011,Bini2015} to the three dimensional (3D) formulations \cite{DeFalco20183D,Bakala2019,Wielgus2019,DeFalco2020,DeFalco2021HT}. The common feature of all models is the presence of the \emph{critical hypersurface}, region where gravitational and radiation forces balance and the test particle moves on it stably \cite{Bini2011,DeFalco2019ST,DeFalco2020sum}. 

From a theoretical point of view, the general relativistic PR effect has been treated under a Lagrangian formalism, determining for the first time in the GR literature the analytical form of the Rayleigh potential \citep{DeFalco2018,DeFalco2019,DeFalco2019VE,DeFalco2020sum}.

There are also several high-energy astrophysical applications, like: investigating the way in which type I X-ray bursts on the NS surface may induce an increased mass inflow rate in the inner edge of the accretion disk \cite{Walker1992,Lancova2017,Fragile2020}; studies on the matter velocity field close to a slowly rotating NS as a result of the PR effect for different star luminosities \cite{Abramowicz1990,Lamb1995,Miller1996,Miller1998}; modelling the photospheric expansion occurring during Eddington-luminosity X-ray bursts \cite{Wielgus2015,Wielgus2016} and associated oscillations \cite{Wielgus2012,Bollimpalli2019}; observational evidences for changes in the inner disk properties, possibly induced by the PR effect from high signal to noise observations of type I X-ray bursts 
\cite{Ballantyne2004,Keek2014a,Keek2014b,Keek2018}; diagnosing the presence of wormholes through the detection of metric-changes occurring in strong field regimes around BHs through the PR critical hypersurfaces \cite{Defalco2020wh,Defalco2021wh};
the PR effect can drive the dynamical evolution of unequal supermassive BHs coalescence in galactic nuclei \cite{Chen2020}.

Recently, it has been shown that the general relativistic PR effect in the equatorial plane of the Kerr metric shows chaotic dynamics for a suitable parameter range \cite{Defalco2021chaos}. In this paper, we would like to explore the timescale of the chaos onset and check how these configurations influence the PR dynamics within the Lyapunov exponents theory.  

The paper is organized as follows: in Sec. \ref{sec:PRmodel} we briefly describe the general relativistic PR effect model and its equations of motion; in Sec. \ref{sec:LEPRE} we calculate the Lyapunov exponents of the general relativistic PR effect; in Sec. \ref{sec:end} we discuss our results proposing some observational classes of astrophysical sources to detect the chaos in the PR dynamics and finally draw the conclusions. 

\section{General relativistic Poynting-Robertson effect in the equatorial plane of the Kerr metric}
\label{sec:PRmodel}
We consider a test particle orbiting in the equatorial plane, $\theta=\pi/2$, around a rotating compact object, whose outside spacetime is described by the Kerr metric. In geometrical units ($c = G = 1$), its line element, ${\rm d}s^2=g_{\alpha\beta}{\rm d}x^\alpha {\rm d}x^\beta$, in standard Boyer-Lindquist coordinates is parametrized by the mass, $M$, and spin, $a$, as 
\begin{equation}\label{kerr_metric}
\begin{aligned}
 \mathrm{d}s^2 &= \left(\frac{2}{r}-1\right)\mathrm{d}t^2 
  - \frac{4a}{r}\mathrm{d}t\, \mathrm{d}\varphi + \frac{r^2}{\Delta}\, \mathrm{d}r^2 
  + \rho \mathrm{d}\varphi^2, 
\end{aligned}
\end{equation}
where $\Delta= r^{2} - 2r + a^{2}$, and $\rho = r^2+a^2+2a^2/r$.

We assume that the motion of the test particle is influenced by the X-ray electromagnetic radiation field coming from an emitting source located in the vicinity of the rotating compact object. Radiation is treated here as a field superimposed on the Kerr spacetime, located in the equatorial plane, and constituted by a coherent flux of photons, propagating outwards from the center of the radiation source. At each time instant the test particle is hit by a photon, moving along null geodesics in the equatorial plane on the Kerr spacetime.
The radiation field includes also the effects of non-zero angular momentum of the photon field, $b$, that would result from the rotation of the central radiation source and/or frame dragging due to the rotating compact object. Therefore, the photons can be emitted either radially, $b=0$, or in any other direction, $b\neq0$, in the equatorial plane. It is useful to introduce the parameter $\beta$, defined as the azimuthal angle of the photon four-momentum measured clockwise from the $\boldsymbol{\varphi}$-axis in the local zero angular momentum observer (ZAMO) frame, which is related to the photon angular momentum $b$ through the formula \citep{Bini2011}
\begin{equation} \label{ANG1}
\cos\beta=\frac{b N}{\sqrt{\rho}(1+b N^\varphi)}, 
\end{equation}
where $\beta\in[0,2\pi]$ and the functions $N,N^\varphi$ read as
\begin{equation}
N=\sqrt{\frac{\Delta}{\rho}},\qquad N^\varphi=-\frac{2a}{r\Delta}.
\end{equation}
It is important to note that for $\sin\beta>0$ ($\sin\beta<0$) we are considering outgoing (ingoing) photons \cite{Bini2011}. 

We assume that the interaction between the radiation field and the test particle occurs through coherent and isotropic scattering in the test particle rest frame with Thomson cross section $\sigma_{\rm T} = 6.7 \times 10^{-25}$~cm$^2$. The relative luminosity of the radiation field is encoded in the parameter $A$, which can be written as $A/M=L_\infty/L_{\rm Edd}$ \citep{Bini2009}, where $L_\infty$ stays for the luminosity measured by an observer at infinity and $L_{\rm Edd}\simeq 1.26 \times 10^{38}\,(M/M_{\odot})$ erg~s$^{-1}$ represents the Eddington limiting luminosity.

The dynamical system describing the equatorial motion of the test particle influenced by the Kerr gravitational field, the radiation pressure and the PR effect is governed by a set of four coupled and fully general relativistic ordinary differential equations of the first order. The first two equations describe the test particle motion in the ZAMO frame in terms of the local spatial velocity $\nu$ and its azimuthal angle $\alpha$ measured clockwise with respect to the $\boldsymbol{\varphi}$-axis in the ZAMO frame. The last two equations transform these ZAMO quantities into the radial $r$ and angular $\varphi$ velocities. The set of equations of motion in the Kerr metric reads as \cite{Bini2009,Bini2011}
\begin{eqnarray}
\label{dnudtau1}
\frac{\rm d \nu}{\rm d t}&=&-\frac{N\sin\alpha}{\gamma^2}[a(n)^{\hat r}+2\nu\cos \alpha\, \theta(n)^{\hat r}{}_{\hat \varphi}]\label{eq:EoM1}\\
&& + A\frac{(1+b N^\varphi)}{\gamma Nr \sqrt{\rho}}
\frac{[\cos(\alpha-\beta) -\nu][1-\nu\cos(\alpha-\beta)]}{|\sin \beta|}, \notag\\
\frac{\rm d \alpha}{\rm d t}&=& -\frac{N\cos \alpha}{\nu}[a(n)^{\hat r} +2\nu\cos \alpha\, \theta(n)^{\hat r}{}_{\hat \varphi} +\nu^2 k_{\rm (lie)}(n)^{\hat r}]\notag  \\
&&+A \frac{  (1+b N^\varphi)}{\gamma Nr \sqrt{\rho}}\frac{[1-\nu\cos(\alpha-\beta)]\sin(\beta -\alpha)}{\nu|\sin \beta|}, \label{eq:EoM2}\\
\frac{\rm d r}{\rm d t}&=&\frac{N\nu\sin\alpha}{\sqrt{g_{rr}}},\label{eq:EoM3}\\
\frac{\rm d \varphi}{\rm d t}&=&\frac{N\nu\cos\alpha}{\sqrt{g_{\varphi\varphi}}}-N^\varphi.\label{eq:EoM4}
\end{eqnarray}
where $\gamma=1/\sqrt{1-\nu^2}$ is the Lorentz factor, and $a(n)^{\hat r},\theta(n)^{\hat r}{}_{\hat \varphi},k_{\rm (lie)}(n)^{\hat r}$ are the acceleration vector, the expansion tensor, and the relative Lie curvature tensor, respectively, whose explicitly expressions are \citep{Bini2011}
\begin{eqnarray}
a(n)^{\hat r}&=&\frac{M[(r^2+a^2)^2-4a^2Mr]}{r^3\rho \sqrt{\Delta}},\\
\theta(n)^{\hat r}{}_{\hat \varphi}&=&\frac{-aM(3r^2+a^2)}{r^3\rho},\\
k_{\rm (lie)}(n)^{\hat r}&=&\frac{- \sqrt{\Delta}(r^3-a^2M)}{r^3\rho}.
\end{eqnarray}

\subsection{Critical hypersurface}
\label{sec:CH}
The general relativistic PR effect model admits as a particular solution, one where gravitational attraction, radiation pressure, and radiation drag force balance. This configuration defines a region, dubbed as \emph{critical hypersurface}, describing a circular orbit in the equatorial plane, where the test particle moves on it stably with constant velocity \cite{Bini2011,DeFalco2019ST}. Therefore, imposing $\nu=\nu_{\rm crit},\alpha=0,r=r_{\rm crit}$ in the equations of motion, we obtain ${\rm d}\nu/{\rm dt}={\rm d}\alpha/{\rm dt}={\rm d}r/{\rm dt}=0$, which entails \cite{Bini2011}
\begin{eqnarray}
&&\nu_{\rm crit}=\cos\beta, \label{eq:CH1}\\
&&a(n)^{\hat r} + 2\theta(n)^{\hat r}{}_{\hat\varphi}\nu_{\rm crit}+k_{\rm (Lie)}(n)^{\rm \hat r}\nu_{\rm crit}^2 \label{eq:CH2}\\
&&=\frac{A(1+bN^\varphi)^2\rm{sgn}(\sin\beta)}{N^2r\sqrt{\rho}\gamma_{\rm crit}^3}.\notag
\end{eqnarray}
The first equation gives the constant velocity of the test particle on the critical hypersurface; whereas the second solved implicitly in terms of $r$ permits to determine the radius of the critical hypersurface in terms of $M,a,A,b$.

\section{Lyapunov exponents of the general relativistic PR effect}
\label{sec:LEPRE}
The theory of the Lyapunov exponents is a generalization of the linear stability theory for dynamical systems, that allows to quantify the rate of separation of infinitesimally close trajectories. Roughly speaking, it relies on studying the asymptotic properties of the tangent map to the dynamical system, and this is usually achieved analysing the linearized flow. We give a short account on the method, referring to \cite{Benettin1,Benettin2} for more details.

We first quickly introduce the general framework, before describing the concrete situation we are interested in. Let $\Phi^t$ be a differentiable flow on an $n$-dimensional connected Riemannian manifold $(\mathcal{M},h)$, where $h$ is the metric on the manifold $\mathcal{M}$\footnote{In Sec. \ref{sec:LT}, we will see that the spacetime $\mathcal{M}=\mathbb{R}^3$ and the metric $h$ on $\mathcal{M}$ is the usual euclidean metric.}, and $d\Phi^t_x:T_x\mathcal{M}\to T_x\mathcal{M}$ the associated tangent map. Here $T_x\mathcal{M}$ denotes the tangent space to $\mathcal{M}$ at $x\in\mathcal{M}$. The Lyapunov exponents are associated with the evolution of tangent vectors to $\mathcal{M}$. Namely, given a non-zero $\boldsymbol{v}\in T_x\mathcal{M}$ they are defined as
\begin{equation}\label{eq:LCE}
\chi(x,\boldsymbol{v}):=\lim_{t\to+\infty}\frac{\ln \Vert d\Phi^t_x\boldsymbol{v}\Vert_h}{t}\,.
\end{equation}
As $\boldsymbol{v}$ varies in $T_x\mathcal{M}$, $\chi(x,\boldsymbol{v})$ takes $s\leq n$ distinct values $\nu_1(x)>\ldots>\nu_s(x)$, with $s=s(x)$. Then one can compute \emph{all} the Lyapunov exponents, letting vary vectors $\boldsymbol{v}\in T_x\mathcal{M}$ in Eq. \eqref{eq:LCE}. However, as explained in \cite{Benettin1}, in practical computations a random choice of vectors always leads to the \emph{largest} Lyapunov exponent, because the others have essentially zero probability to be obtained. 

Let $\chi_i(x)$, with $\chi_1(x)\geq\ldots\chi_n(x)$, be the Lyapunov exponents at $x$ counted with their multiplicity. The sum of the first $p\ge1$ Lyapunov exponents can be computed choosing $p$ vectors $\boldsymbol{v_1},\ldots,\boldsymbol{v_p}\in T_x\mathcal{M}$ and then evaluating the volume of the parallelepiped generated by the transformed vectors $d\Phi^t\boldsymbol{v_1},\ldots,d\Phi^t\boldsymbol{v_p}$ through the formula: 
\begin{equation}\label{eq:pLE}
\lim_{t\to\infty}\frac{1}{t}\operatorname{Vol^p_h}([d\Phi^t_x \boldsymbol{v_1},\ldots,d\Phi^t_x\boldsymbol{v_p}])=\sum^p_{i=1}\chi_i(x)\,.
\end{equation}
As an example of the applicability of the above formula, let us suppose, for instance, we want to compute the second Lyapunov exponent, $\chi_2(x)$. The first one $\chi_1(x)$ is easily determined by  randomly choosing a vector $\boldsymbol{v_1}\in T_x\mathcal{M}$ and then employing Eq. \eqref{eq:LCE}. Therefore, now one needs to choose just another independent vector $\boldsymbol{v_2}\in T_x\mathcal{M}$, compute the volume of the paralleliped $[d\Phi^t\boldsymbol{v_1},d\Phi^t\boldsymbol{v_2}]$, and finally take the limit for $t\to\infty$: 
\begin{equation}
\lim_{t\to\infty}\frac{1}{t}\operatorname{Vol^2_h}([d\Phi^t_x v_1,d\Phi^t_xv_2])=\chi_1(x)+\chi_2(x)\,.
\end{equation}
Subtracting from this sum the value of $\chi_1(x)$, one obtains $\chi_2(x)$. It is now evident how to iteratively apply this method for computing all the Lyapunov exponents $\chi_i(x)$. One simply needs to add each time a new indepedendent vector $\boldsymbol{v_i}\in T_x\mathcal{M}$ and to use Eq. \eqref{eq:pLE}. 

However, as clarified in \cite{Benettin2}, this procedure requires a careful implementation from a numerical point of view, as two main issues arise. First, randomly choosing $\boldsymbol{v}\in T_xM$, $\Vert dT^t_x \boldsymbol{v}\Vert_h$, it exponentially increases as $t\to\infty$. Moreover, when at least two vectors are involved in \eqref{eq:pLE}, the angle between two of them can rapidly become too small to be numerically handled. In \cite{Benettin2} the authors solve all these issues and then provide tests and examples illustrating their computational strategy.

The concrete setting we are considering is those dynamical systems governed by a set of ordinary differential equations of the following type
\begin{equation} \label{eq:DS}
\frac{{\rm d} X_i(t)}{{\rm d} t}=f_i(X_1(t),\dots,X_n(t)),\quad i=1,\dots,n,
\end{equation}
where $f_i$ are supposed to be smooth functions for all $i=1,\dots,n$, and $t$ plays the role of the time. Let us assume that a time-dependent stationary solution $(X_1^*(t),\dots,X_n^*(t))$ exists for the dynamical system (\ref{eq:DS}), i.e., $f_i(X_1^*(t),\dots,X_n^*(t))=0$ for all $i=1,\dots,n$. Let us consider now a small perturbation around the stationary solution, i.e., $X_i(t)=X_i^*(t)+\delta X_i(t)$ for all $i=1,\dots,n$, and linearize thus the dynamical system (\ref{eq:DS}),
\begin{equation} \label{eq:LDS}
\frac{{\rm d} (\delta X_i(t))}{{\rm d} t}=\mathbb{A}_{ij}(t)\delta X_j(t),\quad i,j=1,\dots,n,    
\end{equation}
where $\mathbb{A}_{ij}(t)$ is the \emph{linear stability matrix} obtained as 
\begin{equation} \label{eq:Lmatrix}
\mathbb{A}_{ij}(t)=\left.\frac{\partial f_i}{\partial X_j}\right|_{X_i^*(t)},\quad i,j=1,\dots,n.    
\end{equation}
The solution of the linearized equation (\ref{eq:LDS}) is
\begin{equation}\label{eq:Lsol}
\delta X_i(t)=L_{ij}(t) \delta X_j(0),\quad i,j=1,\dots,n,    
\end{equation}
where the \emph{evolution matrix} $L_{ij}(t)$ satisfies
\begin{equation}\label{eq:Lsol}
\dot{L}_{ij}(t)=\mathbb{A}_{im}(t)L_{mj}(t),\quad i,j=1,\dots,n,
\end{equation}
and $L_{ij}(0)=\delta_{ij}$. We observe that, a priori, $\mathbb{A}_{im}$ depends on the time $t$.
In the special case where the stationary solution is independent from the time $t$, the Lyapunov exponents are simply the eigenvalues of the time-independent matrix $\mathbb{A}_{ij}$. Then the \emph{principal Lyapunov exponent} is the maximal eigenvalue of such a matrix.

\subsubsection{Proprieties of the Lyapunov exponents}
 For \emph{conservative} dynamical systems the sum of all Lyapunov exponents is zero, because a volume element in the phase space is conserved by the flow. Instead if the dynamical system is dissipative the sum of Lyapunov exponents is negative, as the volume element shrinks along a trajectory. Therefore, the Lyapunov exponents provide information concerning local expansion and contraction of phase space, formalizing thus the concept of stretching rate along different directions \cite{Abarbanel1993}. 

Positive Lyapunov exponents are an useful index of the \emph{sensitive dependence on the initial conditions}\footnote{A dynamical system shows sensitive dependence on the initial conditions if tiny perturbations on the initial conditions leads to significantly different future behaviours.} \cite{Wiggins1988,Tabor1989,Strogatz1994,Ott2002,Guckenheimer2002}. Therefore, they are usually taken as a possible indication of chaos provided that some other conditions are satisfied\footnote{The widely accepted definition of \emph{chaos} is due to Robert L. Devaney and it fulfils three proprieties \cite{Devaney2018}: (1) sensitive dependence on initial conditions, (2) topologically mixing (any given region or open set of the phase space eventually overlaps with any other given region in the phase space), (3) presence of a dense set of periodic orbits (every point in the dynamical real space is approached arbitrarily close by periodic orbits).}. Lyapunov exponents are used to characterize unstable orbits, along which chaotic dynamics can develop. As an example,  
the instability of some circular orbits around a Schwarzschild BH can be quantified by a positive principle Lyapunov exponent, although the geodesics around a Schwarzschild BH are not chaotic \cite{Cornish2001}. 
If a dynamical system shows a positive principal Lyapunov exponent, $\bar{\lambda}>0$, it is possible to define the \emph{Lyapunov time} $T_{\bar\lambda}=1/\bar{\lambda}$. This is a characteristic timescale on which a dynamical system is unstable or chaotic and beyond which our predictions break down \cite{Strogatz1994,Cornish2001,Cornish2003}. 

Despite being helpful, Lyapunov exponents must be exploited with caution as they present some drawbacks, especially in GR theory, which can be summarised as:
\begin{itemize}
    \item \emph{they vary from orbit to orbit}, not encoding the collective behaviour of all orbits and not usually catching generic information \cite{Strogatz1994}. In order to get true and appropriate values, they should be averaged over many different points on the same trajectory. Sometimes it can occur that in such a mean process they could return zero values when the considered orbits move in and out of unstable regions \cite{Barrow1981,Barrow1982,Hobill1994};
    \item \emph{they depend on the chosen time coordinate}. Since time is relative in GR, the same happens also for the Lyapunov exponents. If this remark is not taken properly into account can bring to erroneous results. However, whenever a preferred time direction exists, the uncertainty of time can be eliminated. For example, in the Schwarzschild and Kerr spacetimes where a timelike Killing vector exists, the coordinate time of the observer at infinity reveals to be the most appropriate choice \cite{Cornish2001,Cornish2003}.
\end{itemize}

\subsection{Application of the Lyapunov theory to the general relativistic PR effect}
The general relativistic PR effect is \emph{rotationally invariant} (independent from the azimuthal angle $\varphi$) and \emph{autonomous} (does not explicitly depend on the time $t$). Therefore Eqs. (\ref{eq:EoM1}), (\ref{eq:EoM2}), (\ref{eq:EoM3}) represent the dynamical system to investigate. Defined $\boldsymbol{X}=(\nu,\alpha,r)$, the dynamical system can be written as ${\rm d}\boldsymbol{X}/{\rm d}t=\boldsymbol{f}(\boldsymbol{X})$. We linearly perturb it around the critical hypersurface $\boldsymbol{X_0}=(\nu_0,0,r_0)$ (being a stationary solution, i.e., $\boldsymbol{f}(\boldsymbol{X_0})=\boldsymbol{0}$), where $\nu_0$ and $r_0$ can be determined by exploiting Eqs. (\ref{eq:CH1}) and (\ref{eq:CH2}).
We consider the following perturbations
\begin{equation} \label{eq:PERT}
\nu=\nu_0+\varepsilon \nu_1,\quad \alpha=\varepsilon \alpha_1,\quad r=r_0+\varepsilon r_1,\quad \varepsilon\ll1,     
\end{equation}
also written as $\boldsymbol{X}=\boldsymbol{X_0}+\varepsilon \boldsymbol{X_1}$, with $\boldsymbol{X_1}=(\nu_1,\alpha_1,r_1)$. 

The components of the linearized $3\times3$ matrix $\mathbb{A}=({\rm d}\boldsymbol{f}/{\rm d}\boldsymbol{X})_{\boldsymbol{X}=\boldsymbol{X_0}}$ can be found in Table \ref{tab:Table1}, where we obtain the linearized dynamical system ${\rm d}\boldsymbol{X_1}/{\rm d}t=\mathbb{A}\cdot\boldsymbol{X_1}$.

\renewcommand{\arraystretch}{2}
\begin{table*}[ht]
\begin{center}
\caption{\label{tab:Table1} Explicit expressions of the coefficients of the linearized matrix $\mathbb{A}$, where $a_{ij}=\partial f_i/\partial X_j$. The quantities with subscript $0$ mean that they are evaluated in $\nu_0,r_0$. They can be found also in the related \texttt{Mathematica} notebook.}
\normalsize
\begin{tabular}{| c | c |} 
\ChangeRT{1pt}
{\bf Coefficient} & {\bf Explicit expression} \\
\ChangeRT{1pt}
$\chi_1$ & $2 a M \rho ^3 r_0 \left[2 a^2 r_0 \left(-8 M^2+M r_0+6 r_0^2\right)+a^4 (4 M+3 r_0)+r_0^4 (9 r_0-14 M)\right]$\\
\hline
$\chi_2$ & $-3 \rho_0^2 \left[r_0^7 \left(20 a^2 M^2+1\right)-M r_0^6 \left(32 a^2 M^2+5\right)+2 r_0^5 \left(12 a^4 M^2+a^2+3 M^2\right)\right.$\\
& $\left.+a^2 r_0^3 \left(-16 a^2 M^4+4 a^2 M^2+a^2+12 M^2\right)-a^2 M r_0^2 \left(5 a^2+12 M^2\right)+10 a^4 M^2 r_0-8 a^2 M r_0^4-2 a^6 M\right]$\\
\hline
$\chi_3$ & $2 a M \rho_0  r_0^2 \left[-4 \text{Ma}^2 r_0^3 \left(2 a^2 M^2+9\right)+r_0^6 \left(44 a^2 M^2+3\right)-18 M r_0^5 \left(4 a^2 M^2+1\right)+3 r_0^4 \left(16 a^4 M^2+3 a^2+8 M^2\right)\right.$\\
&$\left.+a^2 r_0^2 \left(4 a^2 M^2+9 a^2+36 M^2\right)-2 a^2 M r_0 \left(8 a^4 M^2+9 a^2+12 M^2\right)+3 a^4 \left(a^2+4 M^2\right)\right]$\\
\hline
$\chi_4$ & $r_0 \left[6 a^4 r_0^2 \left(-2 a^4 M^2+12 a^2 M^4+a^2+2 M^2\right)+4 a^2 r_0^4 \left(-8 a^4 M^6-2 a^4 M^2+3 a^2+12 M^2\right)\right.$\\
&$\left.+2 a^2 M r_0^3 \left(16 a^6 M^3+6 a^4 M^2-24 a^2 M^3-15 a^2-4 M^2\right)+3 r_0^8 \left(1-16 a^4 M^4\right)\right.$\\
&$\left.+2 M r_0^7 \left(40 a^4 M^4+6 a^2 M^2-9\right)+2 r_0^6 \left(-24 a^6 M^4-6 a^4 M^2-12 a^2 M^4+5 a^2+18 M^2\right)\right.$\\
&$\left.+2 M r_0^5 \left(8 a^6 M^4+24 a^4 M^2-21 a^2-12 M^2\right)-6 a^6 M r_0 \left(4 a^2 M^2+1\right)+a^8\right]$\\
\hline
$\xi_1$ & $-6 a M r_0^2\rho_0^2 \left(a^2+3 r_0^2\right)$\\
\hline
$\xi_2$ & $\rho_0 \left[2 r_0^5 \left(18 a^2 M^2-1\right)+3 a^2 r_0^3 \left(4 a^2 M^2-1\right)+a^2 r_0 \left(6 M^2-a^2\right)-a^2 M r_0^2-2 a^4 M+3 M r_0^4\right]$\\
\hline
$\xi_3$ & $2 a M r_0 \left[3 r_0^5 \left(1-4 a^2 M^2\right)+a^2 r_0^3 \left(3-4 a^2 M^2\right)+2 a^2 M^2 r_0-a^2 M r_0^2-2 a^4 M-5 M r_0^4\right]$\\
\hline
$\psi_1$ & $\nu_0^2 \left[a^2 \left(2 M^2-M r_0+r_0^2\right)+r_0^3 (r_0-2 M)\right]+M \left(2 a^2 (r_0-2 M)+r_0^3\right)$\\
\hline
$\psi_2$ & $2 a^3 b M r_0^3 \left(-2 M^2+M r_0-3 r_0^2\right)+4 a^5 b M^2 r_0^2+a^2 r_0^4 \left(-12 M^2+2 M r_0+5 r_0^2\right)$\\
& $+a^4 r_0 (2 M+r_0) \left(-6 M^2+M r_0+4 r_0^2\right)+a^6 (2 M+r_0)^2+2 a b M r_0^6 (5 M-3 r_0)+r_0^7 (2 r_0-3 M)$\\
\hline
$\psi_3$ & $-2 a M \nu_0 \left[a^2r_0^3 \left(12 M^2-23 Mr_0+17r_0^2\right)+3 a^6 (M+r_0)-a^4r_0 (M-r_0) (4 M+11r_0)+3r_0^6 (3r_0-7 M)\right]$\\
\hline
$\psi_4$ & $-M \left[2 a^4 r_0 \left(2 M^2+5 M r_0-4 r_0^2\right)+a^2 r_0^4 (25 M-7 r_0)-3 a^6 (M+r_0)-2 r_0^7\right]$\\
\hline
$\psi_5$ & $-a^2 r_0^4 \left(20 M^2-9 M r_0+r_0^2\right)+a^4 M r_0 \left(-2 M^2-5 M r_0+10 r_0^2\right)+3 a^6 M (M+r_0)+r_0^7 (4 M-r_0)$\\
\hline
$a_{11}$ & $\frac{A\gamma_0}{r_0}\frac{\left(1-2 \nu_0^2\right) \left[(\rho_0 -2 a b M r_0)^2+b^2 \Delta_0 \right]-b \sqrt{\Delta_0 } \nu_0 \left(3 \nu_0^2-1\right) (2 a b M r_0-\rho_0 )}{\sqrt{\Delta_0 } \rho_0  (2 a b M r_0-\rho_0 ) \sqrt{ 1-\frac{b^2 \Delta_0 }{(\rho_0 -2 a b M r_0)^2}}}$\\
\hline
$a_{12}$ & $\frac{\frac{A \gamma_0  \sqrt{\rho_0 } r_0^2 \left[\left(\nu_0^2+1\right) (2 a b M r_0+\rho_0 )-2 b \sqrt{\Delta_0 } \nu_0\right]}{\sqrt{\Delta_0 }}-M \left[-4 a^2 M r_0-2 a \sqrt{\Delta_0 } \nu_0 \left(a^2+3 r_0^2\right)+(a^2+r_0^2)^2\right]}{\gamma_0 ^2 \rho_0 ^{3/2} r_0^3}$\\
\hline
$a_{13}$ & $-A\frac{\left\{b \Delta_0 ^{3/2} \left(\nu_0^2+1\right) \left[b^3 \xi_3+b^2 \xi_2+b \xi_1+
\rho_0^3 r_0\left(a^2+3 r_0^2\right)\right]+\nu_0 \left(b^4 \chi_4+b^3 \chi_3+b^2 \chi_2+b \chi_1-r_o\rho^6\right)\right\}}{\gamma_0  \Delta_0 ^{3/2} \rho_0 ^2 r_0^3 \left( 1-\frac{b^2 \Delta_0 }{(\rho_0 -2 a b M r_0)^2}\right)^{3/2}}$\\
\hline
$a_{21}$ & $\frac{\sqrt{\Delta_0} \left(a^4 M+\gamma_0 ^2 r_0 \psi_1\right)-A \gamma_0  \sqrt{\rho_0 } \left(-2 a b M r_0-b \sqrt{\Delta_0 } \nu_0^3+\rho \right)}{\sqrt{\Delta_0 } \rho ^{3/2} r_0^2 \nu_0^2}$\\
\hline
$a_{22}$ & $A\frac{\nu_0 \left[(\rho -2 a b M r_0)^2-2 b^2 \Delta_0 \right]+b \sqrt{\Delta_0 } (\rho_0 -2 a b M r_0)}{\gamma_0  \sqrt{\Delta_0 } \rho_0  r_0 \nu_0 (2 a b M r_0-\rho_0 ) \sqrt{\frac{b^2 \Delta_0 }{(\rho -2 a b M r_0)^2}+1}}$\\
\hline
$a_{23}$ & $-\frac{A \sqrt{\rho_0} r_0\left[\psi_2-b \Delta_0 ^{3/2}r_0^2 \nu_0 \left(a^2+3r_0^2\right)\right]-\gamma_0  \Delta_0  M \left[\sqrt{\Delta_0 } \left(\nu_0^2 \psi_5+\psi_4\right)+\psi_3\right]}{\gamma_0  \Delta_0 ^{3/2} \rho_0 ^{5/2}r_0^5 \nu_0}$\\
\hline
$a_{32}$ & $\frac{\Delta_0\nu_0}{\sqrt{\rho_0} r_0}$\\
\hline
$a_{31}=a_{33}$ & 0\\
\ChangeRT{1pt}
\end{tabular}
\end{center}
\end{table*}
As already explained, since the dynamical system is autonomous the Lyapunov exponents coincide with the eigenvalues of the matrix $\mathbb{A}$, denoted as $\left\{\lambda_1,\lambda_2,\lambda_3\right\}$. The \emph{characteristic eigenvalue equation} in terms of $\lambda$ is 
\begin{equation} \label{eq:CEP}
c_0+c_1\ \lambda+ c_2\ \lambda^2-\lambda^3=0,    
\end{equation}
where
\begin{equation}
\begin{aligned}
&c_0={\rm det}\mathbb{A},\qquad
c_2={\rm Tr}\mathbb{A},\\
&c_1=\sum_{\substack{i,j=1\\ i\neq j}}
^3\left(\frac{a_{ij}a_{ji}-a_{ii}a_{jj}}{2}\right).
\end{aligned}
\end{equation}
Since the eigenvalues are the zeroes of a polynomial of third order, we can analytically determine them \cite{Press2002}. 

\subsubsection{Lyapunov and PR timescales}
\label{sec:LT}
We can calculate the Lyapunov timescale for the chaos onset as $T_{\bar\lambda}=1/\bar{\lambda}$, where $\bar{\lambda}$ is the principal Lyapunov exponent. This time must be compared with the PR timescale $T_{\rm PR}$, defined as the time from the start of the numerical simulation until the test particle reaches for the first time the critical hypersurface for moving then on that stably\footnote{There are some examples, where the test particle can cross the critical hypersurface and then not moving anymore on that (see figures in Sec. 3.4 of Ref. \cite{DeFalcoTESI}, for more details).}. This time is calculated numerically for each simulation. In order to understand whether the chaotic behavior is influential on the global PR dynamics, we compute the ratio $T_{\bar\lambda}/T_{\rm PR}$, and if it is smaller (greater) than one, then the chaos is important (unimportant) \cite{Cornish2003}.

\begin{figure}[h!]
    \centering
    \includegraphics[scale=0.3]{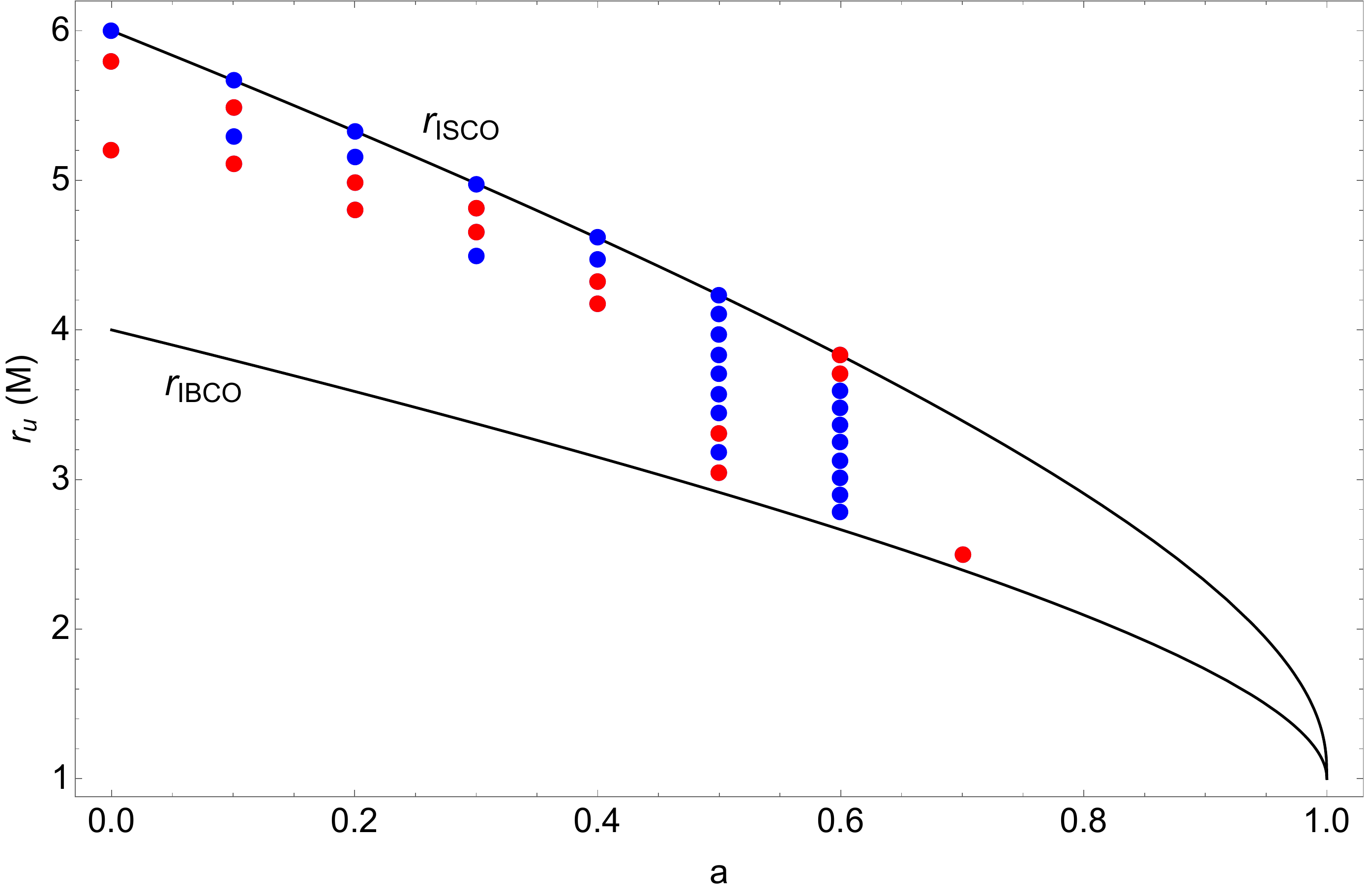}
    \caption{Parameter space $(r_u,a,b)$, where $r_u$ is the homoclinic orbit's parameter, and $b$ is fixed to $b=3$. This space is delimited by the innermost bound circular orbit, $r_{\rm IBCO}$, and the innermost stable circular orbit, $r_{\rm ISCO}$. All the (blue and red) dots represent the parameter subspace in correspondence of which the chaotic dynamic occurs. The red dots are the selected examples reported in simulations 1 -- 15 in Table \ref{tab:Table2}.}
    \label{fig:Fig1}
\end{figure}

As examples we consider selected data from our previous numerical simulations, see Fig. \ref{fig:Fig1} (see also Fig. 7 in Ref. \cite{Defalco2021chaos}, for comparison). The specific data for these simulations are reported in Table \ref{tab:Table2}, see simulations 1 -- 15. We added also other five examples not reported in previous studies in order to show some variability of parameters, see simulations 16 -- 20 in Table \ref{tab:Table2}.


\renewcommand{\arraystretch}{2}
\begin{table*}[ht]
\begin{center}
\caption{\label{tab:Table2} Numerical simulations performed using as test particle's initial conditions $(r_p,\alpha_p,\nu_p)$ and the PR effect model parameters $b,A,a$ together with the critical hypersurface radius $r_0$. As a result we obtain the Lyapunov timescale $T_\lambda$, the PR timescale $T_{\rm PR}$, and the the ratio between these two times in order to see whether the PR effect is important.}
\normalsize
\begin{tabular}{| c || c | c | c || c | c | c | c || c | c | c |} 
\ChangeRT{1pt}
{\bf SIM.}& $\quad\boldsymbol{\nu_p}\quad$  &  $\quad\boldsymbol{\alpha_p}\quad$  &  $\quad\boldsymbol{r_p}\quad$  &  $\quad\boldsymbol{b}\quad$  &  $\quad\boldsymbol{A}\quad$  &  $\quad\boldsymbol{a}\quad$  &  $\quad\boldsymbol{r_0}\quad$  &  $\quad\boldsymbol{T_{\bar{\lambda}}}\quad$  &  $\quad\boldsymbol{T_{\rm PR}}\quad$  &  $\quad\boldsymbol{T_{\bar\lambda}/T_{\rm PR}}\quad$ \\
$\boldsymbol{\#}$ & & & $\boldsymbol{(M)}$ & $\boldsymbol{(M)}$ & & & $\boldsymbol{(M)}$ & $\boldsymbol{(M)}$ & $\boldsymbol{(M)}$ & \\
\hline
\hline
1 & 0.82 & 0.52 & 2.84 & 3 & 0.0001 & 0 & 2.00 & $7.97\times10^6$ & 178703.57 & 44.59 \\
\hline
2 & 0.72 & 0.30 & 3.51 & 3 & 0.0001 & 0 & 2.00 & $7.97\times10^6$ & 178703.57 & 43.60 \\
\hline
3 & 0.89 & 0.76 & 2.39 & 3 & 0.001 & 0.1 & 1.9950 & 587441.09 & 47363.77 & 12.40 \\
\hline
4 &0.92 & 0.91 & 2.25 & 3 & 0.001 & 0.1 & 1.9950 & 587441.09 & 47397.65 & 12.39 \\
\hline
5 &0.79 & 0.45 & 2.80 & 3 & 0.01 & 0.2 & 1.9800 & 14352.60  & 5164.82 & 2.78 \\
\hline
6 &0.75 & 0.35 & 3.03 & 3 & 0.01 & 0.2 & 1.9800 & 14352.60 & 5187.24 & 2.77 \\
\hline
7 &0.75 & 0.34 & 2.92 & 3 & 0.01 & 0.3 & 1.9541 & 2470.75 & 5347.96 & 0.46\\
\hline
8 &0.73 & 0.29 & 3.06 & 3 & 0.01 & 0.3 & 1.9541 & 2470.75 & 5350.97 & 0.46\\
\hline
9 &0.77 & 0.36 & 2.70 & 3 & 0.01 & 0.4 & 1.9167 & 15.79 & 5558.25 & $2.84\times10^{-3}$\\
\hline
10 &0.73 & 0.29 & 2.88 & 3 & 0.01 & 0.4 & 1.9167 & 15.79 & 5583.53 & $2.83\times10^{-3}$\\
\hline
11 &0.29 & 0.73 & 14.31 & 3 & 0.1 & 0.5 & 4.55 & 24.14 & 3397.87 & $7.10\times10^{3}$\\
\hline
12 &0.40 & 0.33 & 7.35 & 3 & 0.1 & 0.5 & 4.55 & 24.14 & 2691.15 & 0.01\\
\hline
13 &0.62 & 0.04 & 3.37 & 3 & 0.1 & 0.6 & 5.62 & 213.32 & 3733.26 & 0.06\\
\hline
14 &0.62 & 0.03 & 3.42 & 3 & 0.1 & 0.6 & 5.62 & 213.32 & 3727.37 & 0.06\\
\hline
15 &0.14 & 0.24 & 21.31 & 3 & 0.1 & 0.7 & 6.16 & 264.20 & 264.20 & 0.06\\
\hline
\hline
16 &0.70 & 0.22 & 3.29 & 3.1 & 0.1 & 0.3 & 1.9698 & 4.83 & 1029.39 & $4.69\times10^{-3}$\\
\hline
17 &0.67 & 0.11 & 3.17 & 3.1 & 0.1 & 0.5 & 5.88 & 619.76 & 4193.14 & 0.15\\
\hline
18 &0.63 & 0.03 & 3.61 & 3.2 & 0.1 & 0.5 & 6.82 & 557.15 & 5227.02 & 0.11\\
\hline
19 &0.61 & 0.05 & 4.03 & 3.3 & 0.1 & 0.4 & 7.22 & 415.77 & 5608.68 & 0.07\\
\hline
20 &0.60 & 0.05 & 4.36 & 3.4 & 0.1 & 0.3 & 7.66 & 499.52 & 6182.17 & 0.08\\
\ChangeRT{1pt}
\ChangeRT{1pt}
\end{tabular}
\end{center}
\end{table*}

\section{Discussions and Conclusions}
\label{sec:end}
The strategy to detect chaos in the PR dynamics relies on computing the \emph{Melnikov integral} $\mathscr{M}$, whose integrating function is the Poisson brackets of the unperturbed Kerr Hamiltonian function and the PR dissipative perturbations. This is a function of mass $M$, spin $a$, homoclinic orbit parameter $r_u$, photon impact parameter $b$, and test particle's initial condition $r_0$, i.e., $\mathscr{M}=\mathscr{M}(M,a,r_u,b;r_0,)$ \cite{Defalco2021chaos}. If there is a combination of these parameters which nullifies $\mathscr{M}$, this means that in correspondence of these values chaotic dynamics occurs. The general relativistic PR effect shows chaos for a suitable range of parameters, see Fig. \ref{fig:Fig1} for an example\footnote{For discovering other ranges of parameters for which chaos occurs, we developed a code written in \texttt{Mathematica}, which permits to facilitate this research (see Ref. [47] in \cite{Defalco2021chaos}, for more details).}.  

As subsequent analysis, in this work we have investigated whether the chaotic dynamics is observationally relevant on the PR dynamics. To this end, we have exploited the theory of Lyapunov exponents, which allows to estimate the timescale of the chaos onset. The procedure is essentially based on perturbing the general relativistic PR equations of motion (\ref{eq:EoM1}) -- (\ref{eq:EoM3}) for low luminosities in terms of the parameter $\varepsilon=A/M\ll1$ around the critical hypersurface values (configuration of equilibrium for the PR dynamical system), see Eq. (\ref{eq:PERT}). Therefore, after performing these calculations, we obtain the linearized matrix $\mathbb{A}$, which is numerically determined once $M,a,A,b,r_0$ have been assigned. Since the PR model is an autonomous dynamical system, the eigenvalues of $\mathbb{A}$ are exactly the \emph{Lyapunov exponents} $\left\{\lambda_1,\lambda_2,\lambda_3\right\}$. Finally, considering the maximum of the real parts of the eigenvalues, we determine the \emph{principal Lyapunov exponent} $\bar{\lambda}$, whose inverse value corresponds to the \emph{Lyapunov timescale} $T_{\bar \lambda}=1/\bar{\lambda}$ for estimating the chaos onset.    
Another fundamental information is encoded in the PR timescale $T_{\rm PR}$, defined as the time from the start of the numerical simulation until the test particle reaches for the first time the critical hypersurface for then moving on it stably. In order to understand how the chaotic behaviour impacts on the PR dynamics, we consider the ratio between the Lyapunov $T_{\bar\lambda}$ and PR $T_{\rm PR}$ timescales: if
$T_{\bar \lambda}/T_{\rm PR}\le1$ it means that the chaos is observationally significant, whereas if $T_{\bar \lambda}/T_{\rm PR}>1$ is unimportant.

We performed 20 numerical simulations, whose detailed values are reported in Table \ref{tab:Table2}. We note that the chaos is significant for simulations 7 -- 20, while there is any influence for simulations 1 -- 6. However, we checked by performing other numerical simulations that varying the luminosity parameter $A$, namely making it smaller and smaller, it is possible to have also for simulations 1 -- 6 a ratio lower than one. Once we calculate the Lyapunov timescale, we have an indication on how to tune the parameters for having a PR timescale such that it is smaller or greater than the time for the chaos onset. We note also that fixed the value of the photon impact parameter $b$, and chosen, for hypothesis, a small value of the luminosity parameter $A$, it follows that an important role is played by the spin $a$. Indeed, the test particle's initial conditions are also fundamental, but they are strongly related to $\left\{M,a,b\right\}$ via the Melnikov integral $\mathscr{M}$ \cite{Defalco2021chaos}. 

From an observational point of view, it is significant to understand how to identify the astrophysical systems where chaos in the PR dynamics can be detected. The requirement of low luminosities permits to provide a first stringent criterion. The mass $M$ and spin $a$ of a compact object can be normally estimated by means of several strategies (see e.g., \cite{Falanga2015,Middleton2016}, for more details). In addition, the surrounding accreting matter can be found distributed almost everywhere with generally different velocities, including some initial configurations for having chaotic dynamics \cite{Frank2002}. Therefore, only the photon impact parameter $b$ must be estimated. It cannot be measured directly from the observational data, but it can be linked to the emitting surface radius $R_\star$ (supposed to be a spherical region) and angular velocity $\Omega_\star$ (assuming that the emitting surface rigidly rotates) through the formula (see Ref. \cite{Bakala2019}, for more details)
\begin{equation} \label{eq:imp_par}
b\equiv\left[-\frac{g_{t\varphi}+g_{\varphi\varphi}\Omega_\star}{g_{tt}+g_{t\varphi}\Omega_\star}\right]_{r=R_\star}=\frac{a^2 R_\star \Omega_\star \rho(R_\star)}{R_\star+2 M (a \Omega_\star -1)}.    
\end{equation}
In addition, we have that $\Omega_\star\in[\Omega_-,\Omega_+]$, being 
\begin{equation}
\Omega_\pm=\frac{-g_{t\varphi}\pm\sqrt{g_{t\varphi}^2-g_{tt}g_{\varphi\varphi}}}{g_{\varphi\varphi}}.    
\end{equation}
Since $R_\star$ can be estimated from the observations, we can relate $\Omega_\star$ in terms of $b,a,M,R_\star$, via Eq. (\ref{eq:imp_par}), as
\begin{equation}
\Omega_\star=\frac{2 a M+b (R_\star-2 M)}{\rho(R_\star)-2M a b}.
\end{equation}
The last condition imposes thus a further constraint to single out the astrophysical systems exhibiting chaos.  

\begin{figure}[h!]
    \centering
    \includegraphics[trim=0cm 1cm 0cm 2cm,scale=0.45]{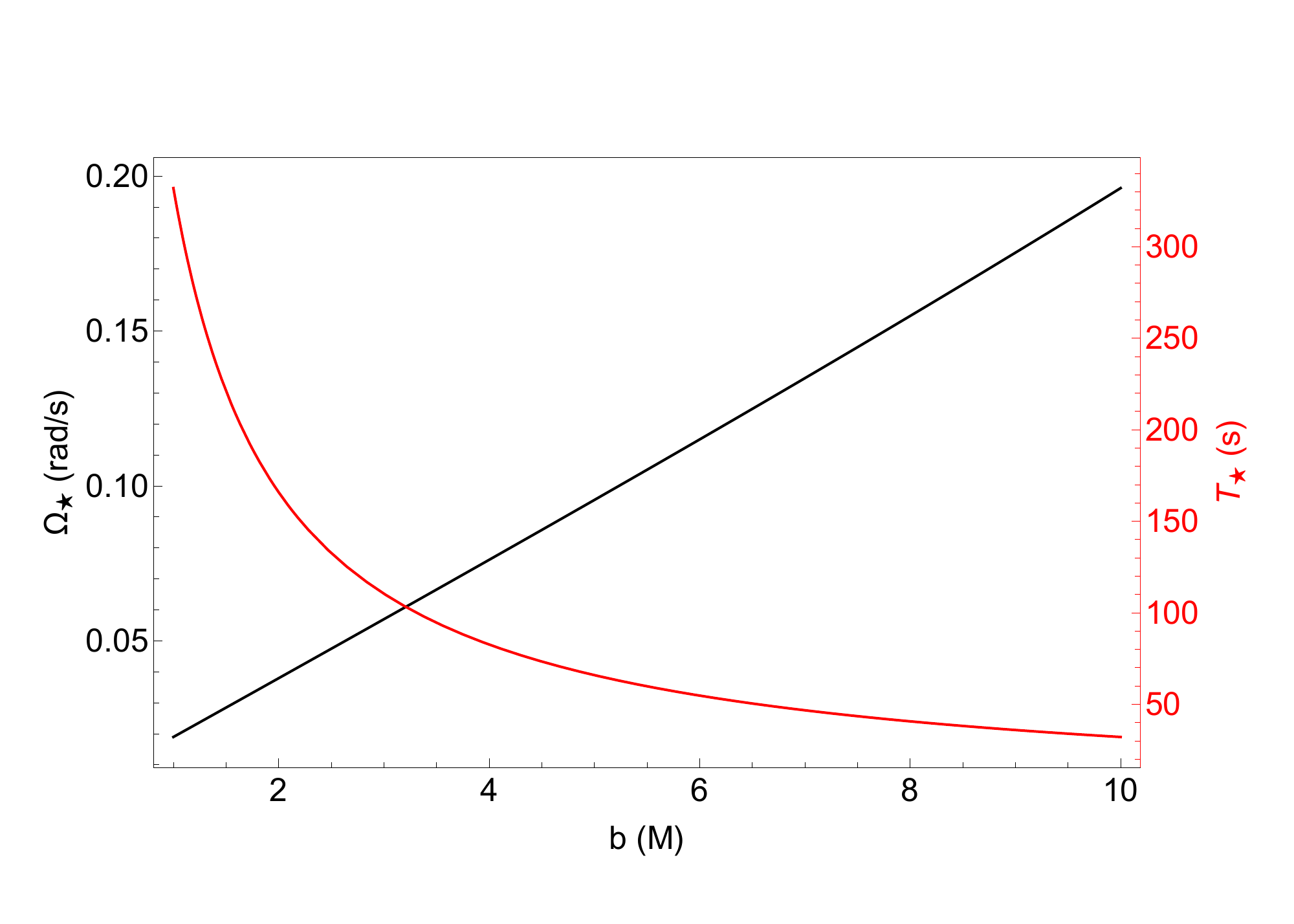}
    \caption{Plot of the NS surface's angular velocity $\Omega_\star$ and spin period $T_\star$ in terms of photon impact parameter $b$.}
    \label{fig:Fig2}
\end{figure}

It is important now to distinguish the physics of BH and NS systems. For a \emph{standard NS} of mass $M=1.4M_\odot$ and radius $R_\star=6M$, if we consider the NS surface as emitting region, then the spin $a$ can be expressed as a function of the NS angular velocity $\Omega_\star$ through $a=\mathcal{C}\Omega_\star/z$, where $\mathcal{C}$ depends on the NS structure and equation of state, which in our case amounts to be $\mathcal{C}=1.1 \times 10^{-4}$ s/rad \cite{Bakala2012}, and $z=(1.4GM_\odot/c^3)/(2\pi)$ is the conversion gravitational factor. In this way, we have $a=a(b)$, which for $b\sim 3$ we obtain $a\sim 10^{-6}$, corresponding thus to extremely slowly rotating NSs, namely $\Omega_\star\sim0.06\ {\rm rad/s}$ or spin period $T_\star\equiv2\pi/\Omega_\star\sim100$ s. In Fig. \ref{fig:Fig2}, we plot how the NS angular velocity $\Omega_\star$ and spin period $T_\star$ changes in terms of the photon impact parameter $b$. In addition, since NSs are very small compact objects, they have a very low luminosity. Assuming that the NS surface temperature is $T=10^6$ K, its luminosity can be calculated through the Stefan-Boltzmann law in terms of the Sun's luminosity (where Sun's temperature $T_\odot=5800$ K and radius $R_\odot=7\times10^5$ Km), obtaining thus $L/L_\odot=0.2$. The Eddington luminosity is $L_{\rm Edd}=4.48\times10^4 L_\odot$, therefore $A\le4.46\times10^-6$. In rows 1, 2, 3 in Table \ref{tab:Table3}, we report the data of some NSs, in which our study can be applied.

\begin{figure}[h!]
    \centering
    \includegraphics[trim=0cm 0cm 0cm 0cm,scale=0.3]{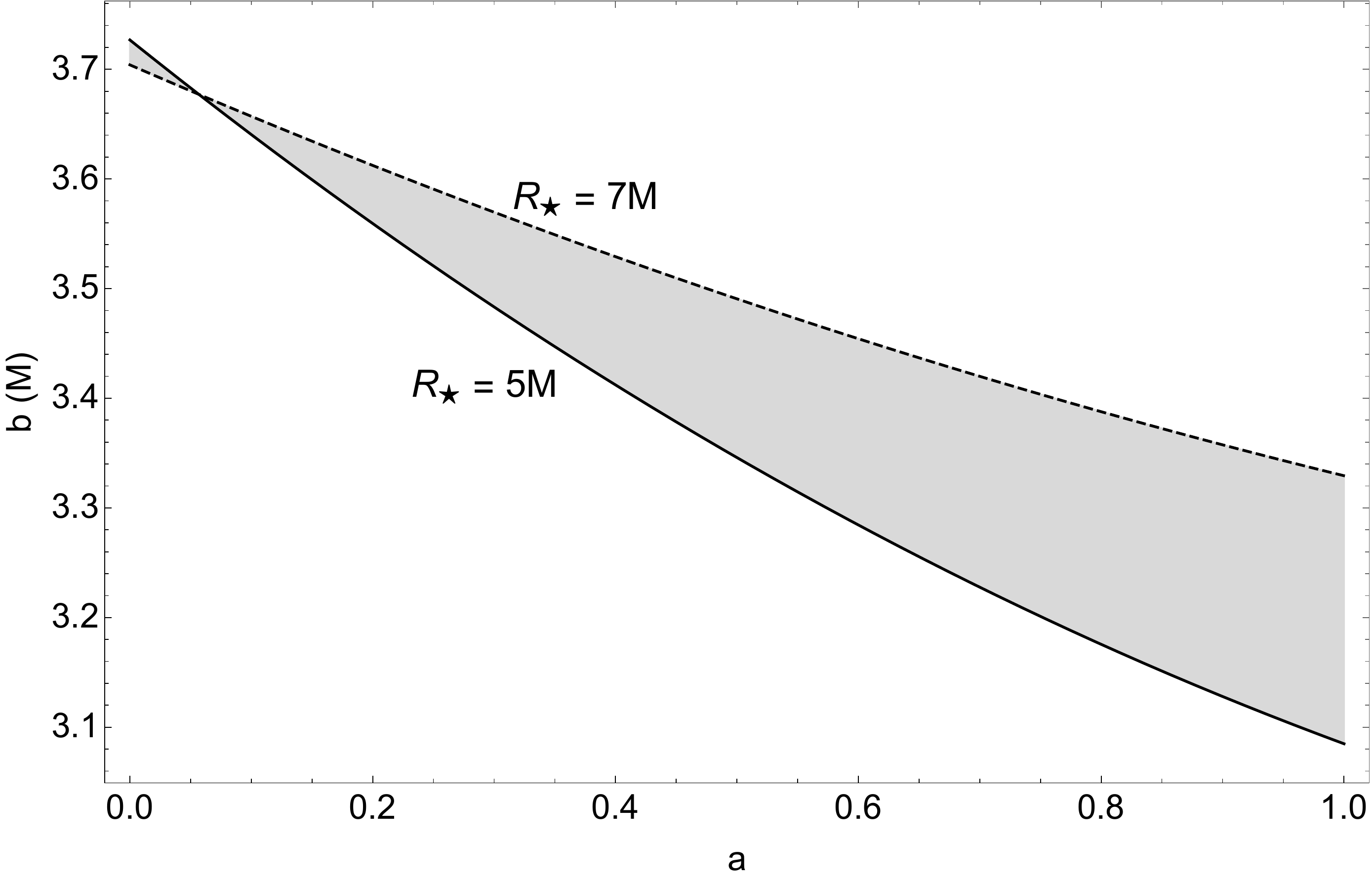}
    \caption{Plot of the NS boundary layer's photon impact parameter $b$ in terms of the NS spin $a$. The continuous ($R_\star=5M$) and dashed ($R_\star=7M$) lines delimit the light grey shaded area, which includes all admissible physical cases.}
    \label{fig:Fig3}
\end{figure}

Instead, if we consider either a boundary layer around a NS or a hot corona around a BH, $a$ and $\Omega_\star$ are now independent. In the case of a \emph{boundary layer around a NS} we know that it is located very close to the NS surface, $R_\star\sim (5 - 7) M$, and it is rigidly rotating with Keplerian angular velocity $\Omega_\star=\Omega_K(R_\star)\equiv M/(a+R_\star)^{3/2}$ \cite{Popham2001}. The additional requirement of low luminosities sees these configurations hosted in atoll sources, characterised by $A\lesssim0.06$ \cite{Ludlam2019}. In Fig. \ref{fig:Fig3} we plot the photon impact parameter $b$ in terms of the NS spin $a$, the only free parameter in this case. Finally in rows 3, 4, 5 in Table \ref{tab:Table3}, there are the data of some astrophysical examples. 

\begin{figure}[h]
    \centering
    \includegraphics[trim=0cm 0cm 0cm 0cm,scale=0.3]{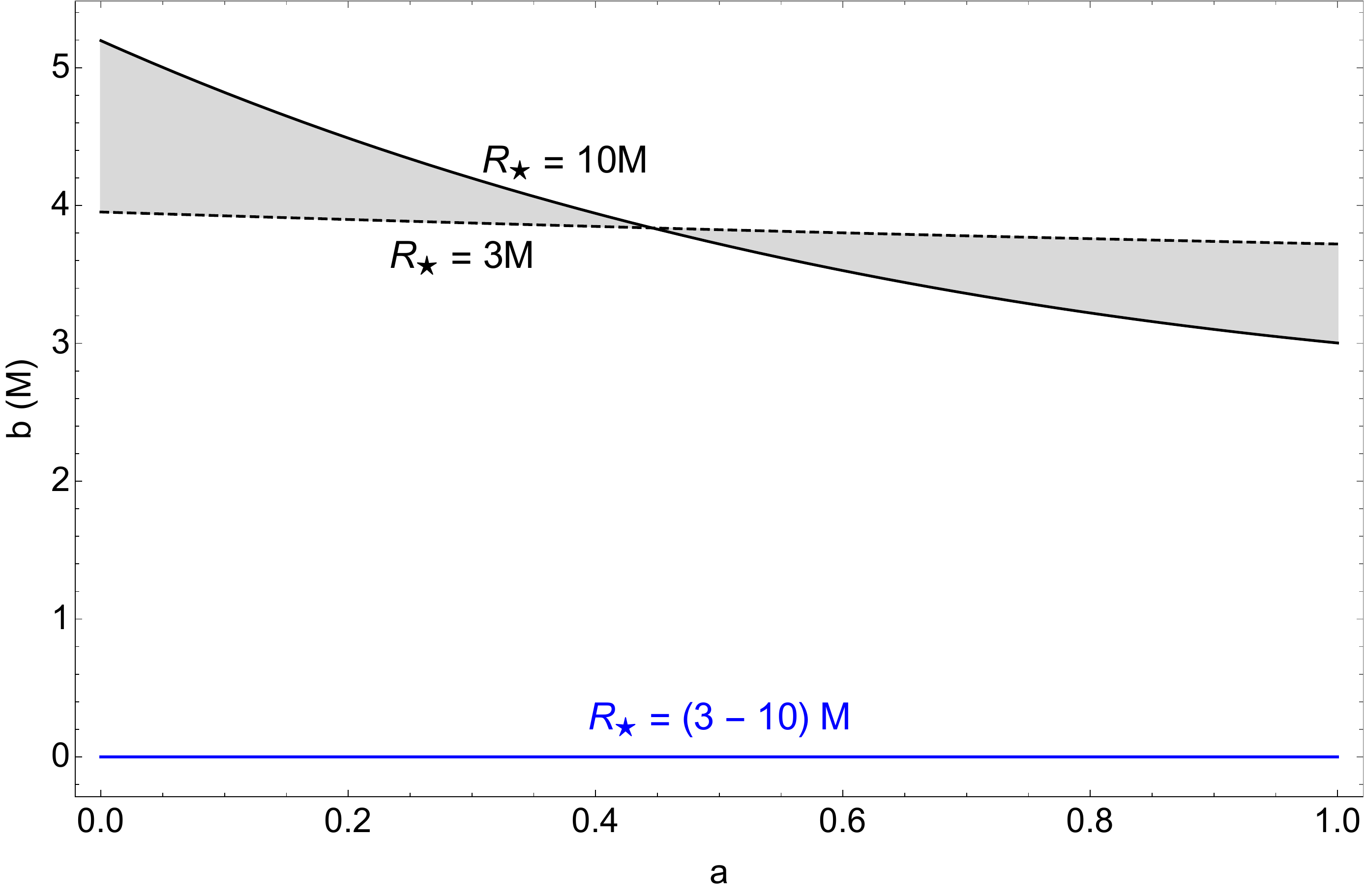}
    \caption{Plot of the BH hot corona's photon impact parameter $b$ in terms of the BH spin $a$ for different locations (light grey shaded area), delimited by $R_\star=3M$ (continuous line) and $R_\star=10M$ (dashed line) in terms of different models of angular velocity $\Omega_\star$: Keplerian $\Omega_K(R_\star)$ coincident with the slab-like in the equatorial plane (black lines), and ZAMO $\Omega_{\rm ZAMO}=[-g_{t\varphi}/g_{\varphi\varphi}]_{r=R_\star}$ (blue line). Looking at Eq. (\ref{eq:imp_par}), it is understandable why to $\Omega_{\rm ZAMO}$ corresponds always $b=0$ for all possible values of $a$ and $R_\star$.}
    \label{fig:Fig4}
\end{figure}

In the BH case, the emitting surface is represented by the hot corona, which is located in the range of $R_\star\sim(3 - 10) M$. We assume that the matter is rotating with Keplerian velocity $\Omega_\star=\Omega_K(R_\star)$. There are however different models of the hot corona's angular velocity proposed in the literature (see Ref. \cite{Niedzwiecki2005}, for more details). We impose also that these sources must be characterised by very small luminosities $A\lesssim0.01$. In Fig. \ref{fig:Fig4} we plot the photon impact parameter $b$ in terms of the BH spin values using different models of the hot corona's angular velocity as reported in Ref. \cite{Niedzwiecki2005}, in order to show how the $b$ range values change in terms of different approaches. In rows 7, 8 ,9 in Table \ref{tab:Table3} we show the data of some astrophysical sources.

The above discussions together with the related formulas, plots, and examples provide some basic strategies, which could be more extensively investigated and improved from an observational point of view through the analysis of several data related to astrophysical BH and NS sources. In addition, our approach is not only restricted to the PR effect, but it can be further extended to study the timescales of other relevant phenomena occurring in high-energy accretion physics (see e.g., \cite{Gaspari2013,King2015,Sukova2016,Blinova2016}).

\renewcommand{\arraystretch}{2}
\begin{table*}[t!]
\begin{center}
\caption{\label{tab:Table3} Information about six examples (ex.) of some astrophysical sources divided in three types analysed in Sec. \ref{sec:end}.}
\normalsize
\begin{tabular}{| c | c | c | c | c | c | c | c |} 
\ChangeRT{1pt}
{\bf Type} & {\bf EX.} & {\bf SOURCE} &$\quad\boldsymbol{M}\quad$ & $\quad\boldsymbol{R_\star}\quad$  &  $\quad\boldsymbol{T_\star}\quad$  &  $\quad\boldsymbol{A}\quad$  &  {\bf Ref.}  \\
& $\boldsymbol{\#}$ & {\bf id.} & $\boldsymbol{{\rm(M_\odot)}}$ &$\boldsymbol{{\rm(M)}}$ & $\boldsymbol{{\rm(s)}}$ & &\\
\hline
\hline
\multirow[c]{3}[4]*{NS surface}& 1 & A0535+26 & 1.50 & 4.52 & 103.00 & $4.46\times10^{-6}$ & \cite{Ikhsanov2001}\\ \cline{2-8}
& 2 & GX 1+4 & 1.35 & 5.03 & 121.00 & $4.46\times10^{-6}$ & \cite{Gonzalez2012}\\ \cline{2-8}
& 3 & Vela X-1 & 1.88 & 3.61 & 283.00 & $4.46\times10^{-6}$ & \cite{Quaintrell2003}\\\cline{2-8}
\hline
\hline
\multirow[c]{3}[4]{1in}{Boundary layer around a NS}& 4 & GX 3+1 & 1.40 & 6.67 & $0.74\times10^{-3}$ & 0.058 & \cite{Ludlam2019}\\ \cline{2-8}
& 5 & 4U 1702--429 & 1.40 & 5.35 & $0.53\times10^{-3}$ & 0.058 & \cite{Ludlam2019}\\ \cline{2-8}
& 6 & GX 301--2 & 1.40 & 5.36 & $0.54\times10^{-3}$ & 0.058 & \cite{Ludlam2019}\\\cline{2-8}
\hline
\hline
\multirow[c]{3}[3]{1in}{Hot corona around a BH}\footnote{The sources we have chosen have all extreme spin values, namely $0.95<a<1$.}& 7 & NGC 5506 & $10^8$ & 10.00 & $1.01\times10^{-3}$ & 0.002 & \cite{Fabian2015}\\ \cline{2-8}
& 8 & MCG-6-30-15 & $0.18\times10^{-3}$ & 2.90  & $0.07$ & 0.03 & \cite{Fabian2015}\\ \cline{2-8}
& 9 & Cyg A & $2.51\times10^9$ & 10.00  & $1.01\times10^{-3}$ & 0.003 & \cite{Fabian2015}\\\cline{2-8}
\ChangeRT{1pt}
\ChangeRT{1pt}
\end{tabular}
\end{center}
\end{table*}

\section*{Acknowledgements}
V.D.F. thanks Gruppo Nazionale di Fisica Matematica of Istituto Nazionale di Alta Matematica for support. W.B. acknowledges support from Gruppo Nazionale per l'Analisi Matematica, la Probabilit\`a e le loro Applicazioni of Istituto Nazionale di Alta Matematica. 

\bibliography{references}
\end{document}